\documentclass[12pt]{article}
\usepackage{graphicx,amsfonts}

\newcommand{\be}{\begin{equation}}
\newcommand{\ee}{\end{equation}}
\newcommand{\bea}{\begin{eqnarray}}
\newcommand{\eea}{\end{eqnarray}}
\newcommand{\beas}{\begin{eqnarray*}}
\newcommand{\eeas}{\end{eqnarray*}}

\begin{document}
\begin{titlepage}

\begin{center}

{\Large Fluid analogs for rotating black holes}

\vspace{10mm}

\renewcommand\thefootnote{\mbox{$\fnsymbol{footnote}$}}
Pablo Garza${}^{1}$\footnote{pablogarza917@gmail.com},
Daniel Kabat${}^{2,3}$\footnote{daniel.kabat@lehman.cuny.edu},
Ariana van Gelder${}^{2,3}$\footnote{avangelder@gradcenter.cuny.edu}

\vspace{4mm}

${}^1${\small \sl Mamaroneck High School} \\
{\small \sl 1000 Boston Post Rd., Mamaroneck NY 10543, USA}

\vspace{2mm}

${}^2${\small \sl Department of Physics and Astronomy} \\
{\small \sl Lehman College, City University of New York} \\
{\small \sl 250 Bedford Park Blvd.\ W, Bronx NY 10468, USA}

\vspace{2mm}

${}^3${\small \sl Graduate School and University Center, City University of New York} \\
{\small \sl  365 Fifth Avenue, New York NY 10016, USA}

\end{center}

\vspace{10mm}

\noindent
Fluid analog models for gravity are based on the idea that any spacetime geometry admits a reinterpretation in which space is thought
of as a fluid flowing with a prescribed velocity.  This fluid picture is a restatement of the ADM decomposition of the metric.
Most of the literature has focused on flat spatial geometries and physical fluid flows, with a view toward possible laboratory realizations.
Here we relax these conditions and consider fluid flows on curved and time-dependent spatial geometries, as a way of understanding and visualizing
solutions to general relativity.  We illustrate the utility of the approach with rotating black holes.  For the Kerr black hole we develop a fluid description
based on Doran coordinates.  For spinning BTZ black holes we develop two different fluid descriptions.  One involves static conical spatial slices, with the fluid orbiting the tip of the cone.  The
other resembles a cosmology, with the fluid flowing on a time-dependent cylindrical geometry.

\end{titlepage}
\setcounter{footnote}{0}
\renewcommand\thefootnote{\mbox{\arabic{footnote}}}

\section{Introduction}

Consider the metric
\be
\label{AcousticMetric}
ds^2 = - c^2 dt^2 + h_{ij} (dx^i - v^i dt) (dx^j - v^j dt)
\ee
where $c(t,x)$, $h_{ij}(t,x)$ and $v^i(t,x)$ are functions of the spacetime coordinates.  Null cones in this metric are determined by
\be
\label{NullCones}
h_{ij} \Big({dx^i \over dt} - v^i\Big) \Big({dx^j \over dt} - v^j\Big) = c^2
\ee
This equation has a simple physical interpretation, in which space is regarded as a fluid moving on a spatial 3-geometry characterized by the line element
\be
\label{SpatialMetric}
ds^2_{\rm spatial} = h_{ij} dx^i dx^j
\ee
In general this spatial geometry could be time-dependent.  The space-fluid is taken to move on this 3-geometry with coordinate velocity $v^i$, and light rays are assumed to
propagate with speed $c$ relative to the moving fluid.  Such light rays obey the null cone condition (\ref{NullCones}), which justifies
the fluid interpretation of the metric (\ref{AcousticMetric}).  We argue this in more detail, and discuss the condition under which fluid flow is geodesic, in appendix \ref{geodesic}.

This interpretation provides an example of analog gravity, a subject which has been explored in the literature beginning with \cite{Unruh:1980cg}.  More recent works include \cite{Visser:2004zs,Natario:2008ej,Ge:2010wx,Hossenfelder:2014gwa,Dey:2016khw,Giacomelli:2017eze}.  The subject has been
reviewed in \cite{Barcelo:2005fc} and gives a perspective which is closely related to the river model of black holes developed in \cite{Hamilton:2004au}.  We should however note an important
distinction.  Much previous work on the subject begins from the equations describing a non-relativistic fluid in flat space and shows that small disturbances propagate according to an effective
curved pseudo-Riemannian metric.\footnote{Below in referring to the literature we will refer to such descriptions as realistic fluid flows.}  In contrast to this previous work we do not impose physical constraints such as
the continuity equation on the motion of the space-fluid, nor do we impose requirements such as a flat Euclidean spatial geometry $h_{ij} = \delta_{ij}$.  Thus
our aim is not to develop fluid models for gravity which could be realized in the laboratory, even in principle.  Instead our aim is to accept the fluid
motion which general relativity prescribes, and to use the analogy as a tool to help understand and visualize properties of solutions.

We are able to do this in complete generality, since (by abandoning considerations of physical fluids and laboratory realizations) any metric can be
put in the so-called acoustic form (\ref{AcousticMetric}).  In fact (\ref{AcousticMetric}) is nothing but the ADM decomposition of the metric, which is
usually presented in the form \cite{Arnowitt:1962hi}
\bea
\nonumber
&& g_{tt} = - (N^2 - h_{ij} N^i N^j) \\
\label{ADM}
&& g_{ti} = g_{it} = h_{ij} N^j \\
\nonumber
&& g_{ij} = h_{ij}
\eea
Thus the speed of light $c$ can be identified with the ADM lapse $N$, and the fluid velocity $v^i = - N^i$ with the negative of the ADM shift.  Incidentally this provides a purely
three-dimensional interpretation of general relativity: rather than describing a foliation of a four-dimensional spacetime, one can think of the lapse and shift as describing a fluid
flow on a spatial three-manifold.

Rather remarkably, several well-known solutions to general relativity have simple fluid analogs.  One prime example is the Schwarzschild
metric in Painlev\'e - Gullstrand coordinates \cite{Painleve,Gullstrand,Kraus:1994fh}
\be
\label{PG}
ds^2 = - dt^2 + \Big(dr + \sqrt{2 G M \over r} dt\Big)^2 + r^2 d\theta^2 + r^2 \sin^2\theta d\phi^2
\ee
Here the spatial geometry is flat and the speed of light $c = 1$.  The fluid moves radially inward, with a velocity
\[
v^r = - \sqrt{2 G M \over r}
\]
This fluid analog very naturally captures the image of space flowing into the singularity at $r = 0$.  The radial velocity of the fluid
reaches the speed of light at the horizon $r = 2 G M$.  The coordinates (\ref{PG}) cover the right exterior and future interior of the black hole, as shown for
example in \cite{Kraus:1994fh}.  Note that with respect to the Killing vector ${\partial \over \partial t}$ the
4-geometry is stationary but not static due to the off-diagonal $g_{tr}$ components of the metric, or equivalently in our language due to the steady
motion of the fluid.

Another example is a flat $k = 0$ FRW cosmology with
\be
\label{FRW1}
ds^2 = - dt^2 + a^2(t) \big(dr^2 + r^2 d\Omega^2\big)
\ee
In terms of the proper radial distance $R = a(t) r$ we have
\be
\label{FRW2}
ds^2 = - dt^2 + (dR - H R \, dt)^2 + R^2 d\Omega^2
\ee
The spatial slices are static and flat, with metric $ds^2_{\rm spatial} = dR^2 + R^2 d\Omega^2$, and the fluid moves radially outward with the Hubble velocity
\[
v^R = H R, \qquad H = {\dot{a} \over a}
\]
Again the speed of light $c = 1$.  Note that the fluid velocity reaches the speed of light at the Hubble or apparent horizon $R = 1/H$.

One of the aims of the present work is to explore fluid analogs for other exact solutions of general relativity, in particular for rotating black holes.  We explore fluid analogs for
the Kerr solution in section \ref{sect:Kerr} and for the BTZ solution in section \ref{sect:BTZ}.  We conclude in section \ref{sect:conclusions}.

\section{Kerr metric\label{sect:Kerr}}

In developing a fluid analog for the Kerr geometry \cite{Kerr:1963ud} the first question which arises is the choice of coordinates.  This is largely a matter of taste,
as any metric can be decomposed in the ADM form (\ref{ADM}).  However particular choices may make the fluid analog particularly simple or appealing.
This was certainly the case for the Schwarzschild and FRW metrics (\ref{PG}), (\ref{FRW2}).  For Kerr the choice of coordinates is less compelling.

One obstacle is that (unlike Schwarzschild) there is no slicing of the Kerr geometry that is metrically or even conformally flat \cite{Garat:2000pn,ValienteKroon:2003ux}.  This led some previous studies to focus on
developing an analog model for the equatorial plane \cite{Visser:2004zs}.
On the equatorial plane one can
develop a fluid analog based on Boyer - Lindquist
coordinates, as shown in \cite{Visser:2004zs}.  However the resulting fluid motion has some rather counter-intuitive features: the motion is purely
angular, and the horizon manifests itself as a locus where the speed of light $c \rightarrow 0$.

We will instead develop a fluid analog based on Doran coordinates \cite{Doran:1999gb}, which were specifically developed to extend the
Painlev\'e - Gullstrand coordinates (\ref{PG}) to rotating black holes.  A similar description was developed in \cite{Natario:2008ej} using somewhat different coordinates.
A realistic analog description for Kerr is an open problem; recent work can be found in \cite{Giacomelli:2017eze,Liberati:2018osj}.  The general problem of constructing coordinates which are regular at the horizon has been
considered in \cite{Zaslavskii:2018lbb}.

In Doran coordinates the Kerr metric is\footnote{Our discussion of the geometry closely follows \cite{Doran:1999gb}.
Also see \cite{Hawking:1973uf,Visser:2007fj}.}\footnote{We set $G = 1$ so the mass $m$ and angular
momentum per unit mass $a = J / m$ have units of length.  We restrict to $-m < a < m$ so the singularity is inside the horizon.}
\bea
\nonumber
ds^2 & = & - dt^2 + {r^2 + a^2 \cos^2 \theta \over r^2 + a^2} \left[dr + {\sqrt{2 m r (r^2 + a^2)} \over r^2 + a^2 \cos^2 \theta} \left(dt - a \sin^2 \theta d\phi\right)\right]^2 \\[3pt]
\label{DoranKerr}
& & + (r^2 + a^2 \cos^2 \theta) d\theta^2 + (r^2 + a^2) \sin^2\theta d\phi^2
\eea
with horizons at $r^2 + a^2 = 2mr$ and ergospheres at $r^2 + a^2 \cos^2 \theta = 2 m r$.  The relation between Doran and Boyer - Lindquist coordinates is given in
appendix \ref{BL-D}.

It is straightforward to extract the analog fluid motion.  Restricting to constant time slices we find that the spatial geometry is static,
\bea
\nonumber
ds^2_{\rm spatial} & = & {r^2 + a^2 \cos^2 \theta \over r^2 + a^2} \left[dr - {a \sin^2 \theta \sqrt{2 m r (r^2 + a^2)} \over r^2 + a^2 \cos^2 \theta} d\phi\right]^2 \\[3pt]
\label{KerrSpatial}
& & + (r^2 + a^2 \cos^2 \theta) d\theta^2 + (r^2 + a^2) \sin^2 \theta d\phi^2
\eea
The fluid moves with coordinate velocity
\bea
\label{KerrVr}
&& v^r = - {\sqrt{2 m r(r^2 + a^2)} \over r^2 + a^2 \cos^2 \theta} \\[5pt]
\nonumber
&& v^\theta = v^\phi = 0
\eea
and the speed of light $c = 1$.  Note that the coordinate velocity of the fluid is purely radial, but this is a bit deceptive because the spatial coordinates $r$ and $\phi$
aren't orthogonal.  An appealing feature of these coordinates is that as $a \rightarrow 0$ we smoothly recover the Schwarzschild expressions in Painlev\'e - Gullstrand
coordinates.

\begin{figure}
\centerline{\includegraphics[width=6.6cm]{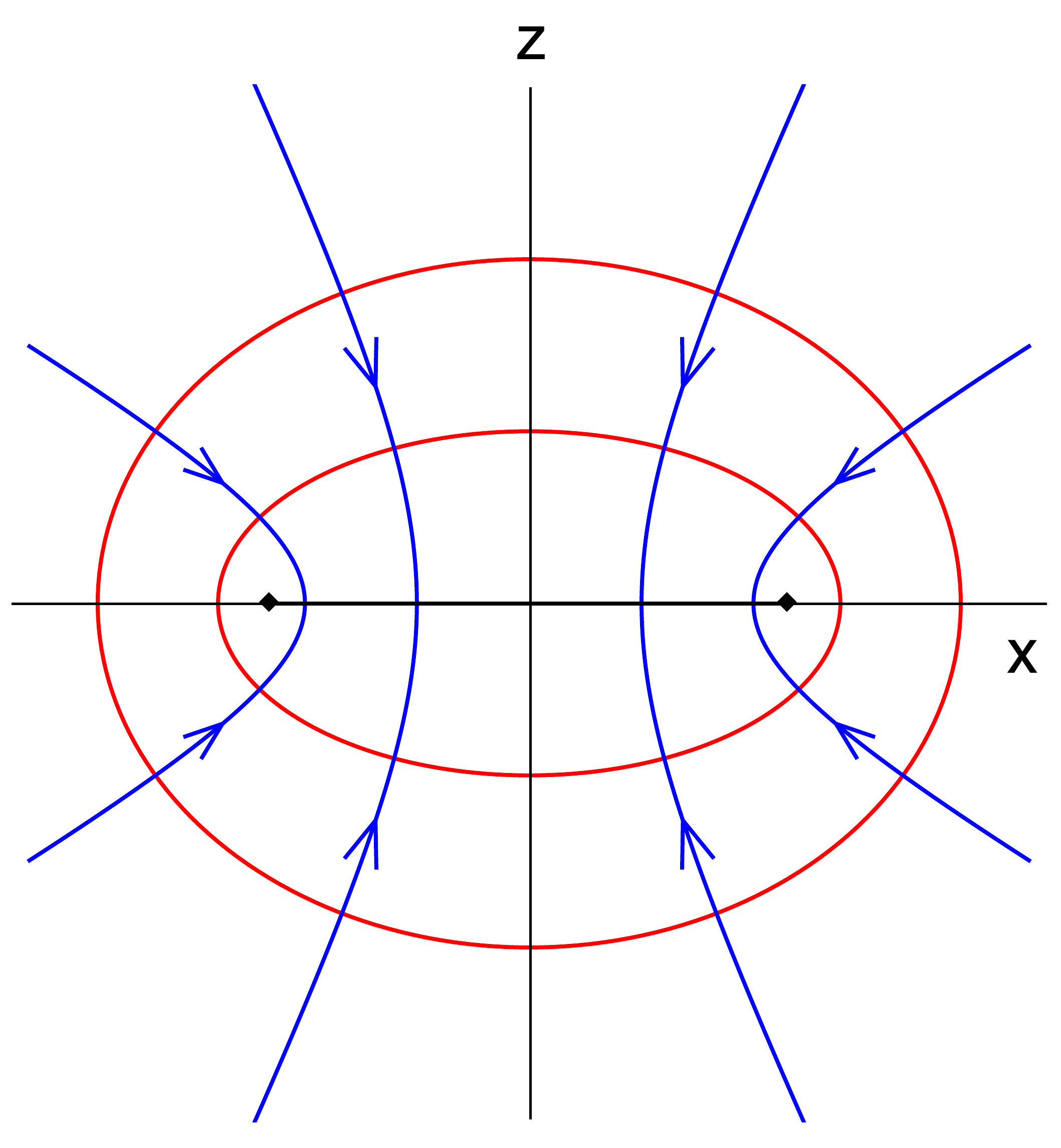}}
\caption{The $(x,z)$ plane has a flat geometry.  The singularity is at $x = \pm a$, $z = 0$.  Surfaces of constant $r$
are ellipses ${x^2 / (r^2 + a^2)} + {z^2 / r^2} = 1$ with foci at the singularity.  The fluid travels toward $r = 0$ along lines of constant $\theta$ which are hyperbolas ${x^2 / \sin^2 \theta} - {z^2 / \cos^2 \theta} = a^2$.\label{KerrFig1}}
\end{figure}

To describe the spatial geometry it's convenient to expand
\bea
\nonumber
ds^2_{\rm spatial} & = & {r^2 + a^2 \cos^2 \theta \over r^2 + a^2} dr^2 + (r^2 + a^2 \cos^2 \theta) d\theta^2 + (r^2 + a^2) \sin^2 \theta d\phi^2 \\
\label{KerrSpatial2}
& & - \left({8mr \over r^2 + a^2}\right)^{1/2} a \sin^2 \theta dr d\phi + {2 m r a^2 \sin^4 \theta \over r^2 + a^2 \cos^2 \theta} d\phi^2
\eea
Here
\be
0 < r < \infty \qquad 0 < \theta < \pi \qquad \phi \approx \phi + 2 \pi
\ee
The first line in (\ref{KerrSpatial2}) is flat space in oblate spheriodal coordinates, related to the usual Cartesian coordinates by
\bea
\nonumber
&& x = \sqrt{r^2 + a^2} \sin \theta \cos \phi \\
&& y = \sqrt{r^2 + a^2} \sin \theta \sin \phi \\
\nonumber
&& z = r \cos \theta
\eea
This is drawn in the $(x,z)$ plane in Fig.\ \ref{KerrFig1}.  Note that the intrinsic geometry of the $(x,z)$ plane is flat.  However the orbits of the Killing vector
${\partial \over \partial \phi}$ aren't orthogonal to the $(x,z)$ plane due to the off-diagonal metric component
\[
h_{r \phi} = - \left({2mr \over r^2 + a^2}\right)^{1/2} a \sin^2 \theta
\]
Also the radius $R$ of the Killing orbits is distorted, from the value it would have in flat space
\[
R^2 = x^2 + y^2 = \big(r^2 + a^2\big) \sin^2 \theta
\]
to
\[
R^2 = x^2 + y^2 + {2 m r a^2 \sin^4 \theta \over r^2 + a^2 \cos^2 \theta}
\]
The singularity is a ring at $r = 0$, $\theta = \pi/2$ or equivalently $x^2 + y^2 = a^2$, $z = 0$.  Note that the proper distance around the singularity depends on how it is
approached.  If one approaches from the inside in the equatorial plane ($r \rightarrow 0$ then $\theta \rightarrow \pi/2$) ring has radius $a$, but if one approaches from
the outside ($\theta \rightarrow \pi/2$ then $r \rightarrow 0$) the radius diverges.  Also note that the fluid travels towards $r = 0$ -- that is, towards
the disk $x^2 + y^2 < a^2$, $z = 0$ -- along lines of constant $\theta$, $\phi$.  We will not consider the extension of the solution to $r < 0$.

One property of the fluid motion is easy to understand: the fluid reaches the speed of light at the ergosphere.
\bea
&& \Vert v \Vert^2 = h_{ij} v^i v^j = {2 m r \over r^2 + a^2 \cos^2 \theta} \\[5pt]
\nonumber
&& r^2 + a^2 \cos^2 \theta = 2 m r \quad \Rightarrow \quad \Vert v \Vert = 1
\eea
Since the fluid is moving at the speed of light at this radius, it's clear this locus defines the stationary limit surface.
To understand the horizon in these terms we decompose the fluid velocity
\be
v^i = v_{\rm radial}^i + v_{\rm angular}^i
\ee
into orthogonal components in the radial and angular directions.
\bea
\nonumber
&& v_{\rm radial} = v^r \partial_r - u^\phi \partial_\phi \\
&& v_{\rm angular} = u^\phi \partial_\phi
\eea
Here
\be
u^\phi = {2 m r a \over (r^2 + a^2) (r^2 + a^2 \cos^2 \theta) + 2 m r a^2 \sin^2 \theta}
\ee
is chosen so that $\langle v_{\rm radial}, v_{\rm angular} \rangle = 0$.  Then the magnitude of the radial velocity satisfies
\be
\Vert v_{\rm radial} \Vert^2  = {2 m r (r^2 + a^2) \over (r^2 + a^2) (r^2 + a^2 \cos^2 \theta) + 2 m r a^2 \sin^2 \theta}
\ee
and the radial velocity reaches the speed of light at the horizon $r^2 + a^2 = 2mr$.  Since the radial velocity reaches the
speed of light, it's clear this locus defines a trapped surface.

\subsection{Equatorial Kerr}

The spatial 3-geometry of the Kerr metric (\ref{KerrSpatial}) is somewhat complicated by the presence of the off-diagonal $h_{r\phi}$ components of the metric.
As a simple setting where this can be avoided we focus on the equatorial plane of the Kerr geometry, following \cite{Visser:2004zs}.  After setting
$\theta = \pi / 2$ in (\ref{DoranKerr}) the spatial coordinates can be ``straightened'' by setting
\be
\phi = \tilde{\phi} + f(r)
\ee
with
\be
\left({r^2 + a^2 \over a} + {2 m a \over r}\right)f'(r) = \sqrt{2 m r \over r^2 + a^2}
\ee
This leads to
\bea
\nonumber
ds^2 & = & - \left(1 - {2 m \over r}\right) dt^2 + {\sqrt{8 m r(r^2 + a^2)} \over r^2 + a^2 + {2 m a^2 \over r}} dt dr - {4 m a \over r} dt d\tilde{\phi} \\[5pt]
& & \quad + {r^2 dr^2 \over r^2 + a^2 + {2 m a^2 \over r}} + \left(r^2 + a^2 + {2 m a^2 \over r}\right) d\tilde{\phi}^2
\eea
One then reads off
\bea
\nonumber
&& \hbox{\rm spatial metric} \quad ds^2_{\rm spatial} =  {r^2 dr^2 \over r^2 + a^2 + {2 m a^2 \over r}} + \left(r^2 + a^2 + {2 m a^2 \over r}\right) d\tilde{\phi}^2 \\[5pt]
\nonumber
&& \hbox{\rm fluid velocity} \quad v^r = - \sqrt{2 m (r^2 + a^2) \over r^3} \\[5pt]
&& \hspace{28mm} v^{\tilde{\phi}} = {2 m a \over r(r^2 + a^2) + 2 m a^2} \\[5pt]
\nonumber
&& \hbox{\rm speed of light} \quad c = 1
\eea

\begin{figure}
\centerline{\includegraphics[width=8.1cm,height=8.1cm]{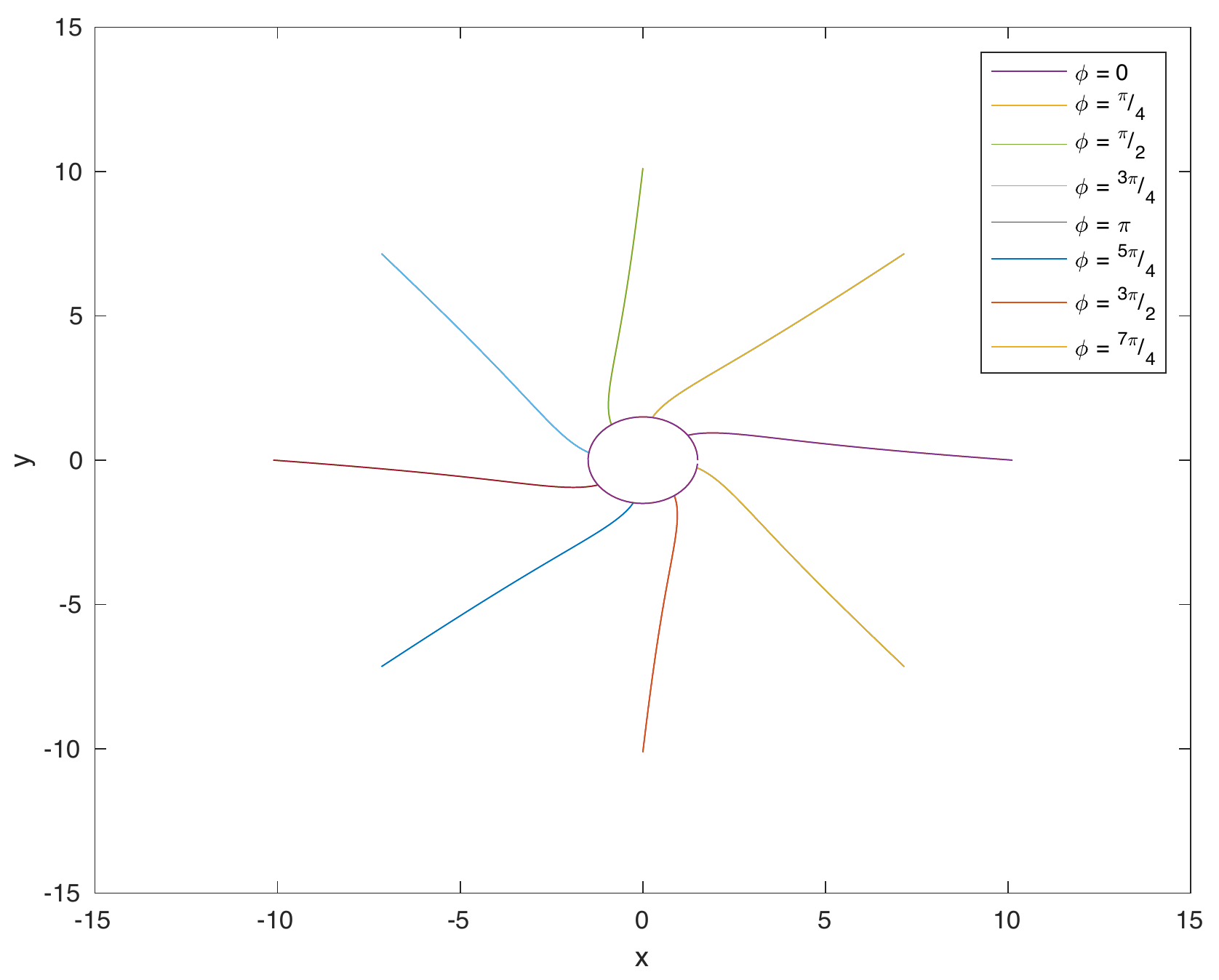}}
\caption{Fluid streamlines in the Kerr equatorial plane.  The axes are $x = \sqrt{r^2 + a^2} \cos \tilde{\phi}$, $y = \sqrt{r^2 + a^2} \sin \tilde \phi$.\label{KerrFig2}}
\end{figure}

Some fluid streamlines in the equatorial plane are shown in Fig.\ \ref{KerrFig2}.  As the fluid moves from $r = \infty$ to the singularity it rotates by an angle
\be
\Delta \tilde{\phi} = \int_0^\infty dr \, {v^{\tilde{\phi}} \over v^r}
\ee
For a non-rotating black hole of course $\Delta \tilde{\phi} = 0$, and in the extremal limit $a = \pm m$ this gives $\Delta \tilde{\phi} \approx \pm 36^\circ$.

\section{BTZ black hole\label{sect:BTZ}}

\begin{figure}[h]
\centerline{\includegraphics[width=6.5cm]{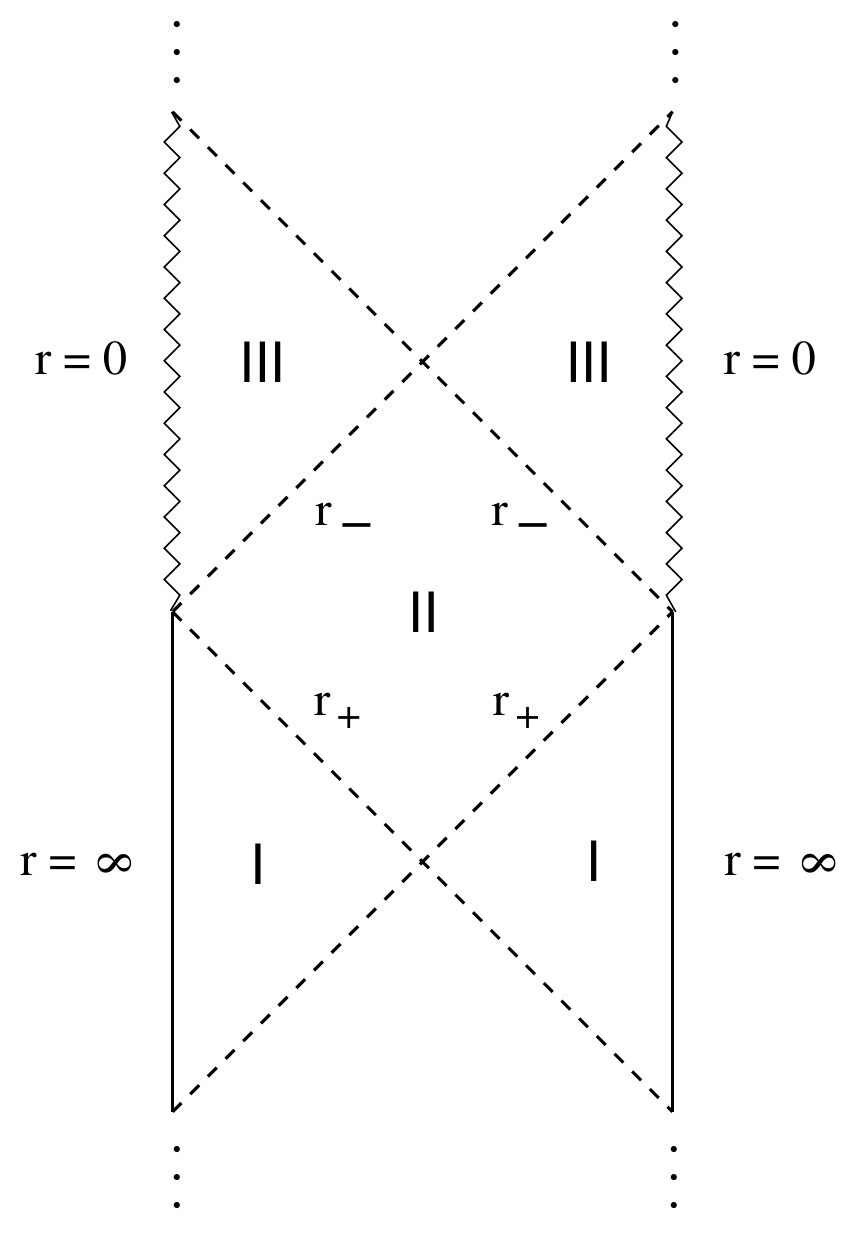}}
\caption{Penrose diagram for a spinning BTZ black hole.  The cosmological coordinates of section \ref{sect:Spinning_BTZ} cover region {\sf II}.\label{fig:Spinning_BTZ}}
\end{figure}

\begin{figure}
\centerline{\includegraphics[width=6cm]{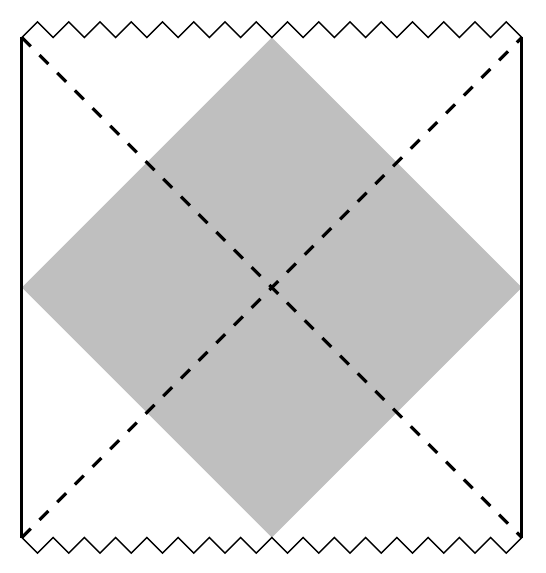}}
\caption{Penrose diagram for a BTZ black hole with $J = 0$.  The cosmological coordinates (\ref{CosmologicalBTZ}) cover the shaded diamond.\label{fig:BTZ}}
\end{figure}

In this section we consider the BTZ black hole \cite{Banados:1992wn,Carlip:1995qv}.  In Schwarzschild coordinates the metric is
\bea
\label{BTZ-Schwarzschild}
ds^2 & = & - f(r) dt^2 + {1 \over f(r)} dr^2 + r^2 \left(d\phi - {r_+ r_- \over \ell r^2} dt\right)^2 \\[5pt]
\nonumber
f(r) & = & {(r^2 - r_+^2)(r^2 - r_-^2) \over \ell^2 r^2}
\eea
The coordinate $\phi$ is periodically identified with period $2\pi$.  The geometry is asymptotic to AdS${}_3$ with radius of curvature $\ell$.  Inner and outer horizons
are located at $r = r_-$ and $r = r_+$, respectively.  These parameters are related to the mass and angular momentum of the black hole by\footnote{in units where $8G = 1$}
\be
M = {1 \over \ell^2} \big(r_+^2 + r_-^2\big) \qquad\qquad J = {2 r_+ r_- \over \ell}
\ee
The Penrose diagram for a spinning black hole is shown in Fig.\ \ref{fig:Spinning_BTZ} while the spinless case is shown in Fig.\ \ref{fig:BTZ}.  A realistic fluid description of BTZ
was developed in \cite{Giacomelli:2017eze}.  In this section we consider some instructive generalizations.

\subsection{Conical acoustic metric\label{sect:ConicalAcoustic}}

We first consider ``conical coordinates'' which lead to simple static spatial slices.

The BTZ metric is invariant under shifts of $t$ and $\phi$, so it's natural to consider reparametrizing
\be
\label{BTZshift}
t \rightarrow t + g(r) \qquad\quad \phi \rightarrow \phi + h(r)
\ee
in terms of two arbitrary functions $g(r)$ and $h(r)$.
Setting $h' = {r_+ r_- \over \ell r^2} g'$ eliminates the off-diagonal components of the resulting spatial metric and puts the BTZ metric in the form
\be
ds^2 = - f(r) \left(dt + g'(r) dr\right)^2 + {1 \over f(r)} dr^2 + r^2 \left(d\phi - {r_+ r_- \over \ell r^2} dt\right)^2
\ee
At this stage it's convenient to choose a constant $c$ which is positive and dimensionless but otherwise arbitrary.
(The name is not misleading as $c$ will shortly be identified with the speed of light.)
Setting ${1 \over f} - f g'{}^2 = {1 \over c^2}$ brings the metric to a conical acoustic form with
\bea
\label{SpatialCone}
&& \hbox{\rm spatial metric} \quad ds^2_{\rm spatial} =  {1 \over c^2} \left(dr^2 + c^2 r^2 d\phi^2\right) \\[5pt]
\label{BTZvr}
&& \hbox{\rm fluid velocity} \quad v^r = \pm c\sqrt{c^2 - f(r)} \\[5pt]
\nonumber
&& \hspace{28mm} v^{\phi} = {r_+ r_- \over \ell r^2} \\[5pt]
\nonumber
&& \hbox{\rm speed of light} = c
\eea
This generalizes the results of \cite{Giacomelli:2017eze} who considered the case $c = 1$.
Note that the choice of $\pm$ in (\ref{BTZvr}) corresponds to an outgoing or ingoing fluid, that is, to a white or black hole patch of the geometry.

The spatial geometry (\ref{SpatialCone}) is a cone with a total angle $2 \pi c$ about the tip.  The fluid flow (\ref{BTZvr}) is rather curious as there are
radial turning points where $f(r) \rightarrow c^2$ and $v^r \rightarrow 0$.  So the flow is bounded between the radii $r_{\rm min}$ and $r_{\rm max}$ where
\be
\big(r_{{\rm max} \atop {\rm min}}\big)^2 = {1 \over 2} \left(c^2 \ell^2 + r_+^2 + r_-^2 \pm \sqrt{\left(c^2 \ell^2 + r_+^2 + r_-^2\right)^2 - 4 r_+^2 r_-^2}\right)
\ee
Note that $r_{\rm min} < r_- < r_+ < r_{\rm max}$.  The intuitive picture is that the asymptotic AdS geometry acts as a gravitational potential well
which pushes the fluid back toward the black hole.  This accounts for the outer turning point.  But the singularity of a spinning BTZ black hole is repulsive,\footnote{Note that $g_{tt}$ diverges both as $r \rightarrow 0$ and as $r \rightarrow \infty$.} so the fluid is also pushed away from $r = 0$ which gives rise to the inner turning point.  Thus the fluid spirals up the Penrose diagram of Fig.\ \ref{fig:Spinning_BTZ}, emerging
from a white hole and falling back into the next black hole.  The case $J = 0$ is special in that there is no inner turning point.  The fluid emerges from the white hole,
reaches a maximum radius, then falls back into the black hole.

From a spacetime perspective the arbitrary constant $c$ we have introduced controls the way that constant-time slices are embedded in the $(2+1)$-dimensional geometry.  In the fluid interpretation $c$
plays the role of the speed of light with respect to the fluid.  Somewhat curiously $c$ also controls the deficit angle of the spatial cone, with
$c = 1$ corresponding
to flat spatial slices.  Finally the radial fluid velocity $v^r$ increases with $c$, which means the range of radii covered by these coordinates $r_{\rm min} < r < r_{\rm max}$
grows with $c$.  The limit $c \rightarrow \infty$ is however singular.

\begin{figure}
\centerline{\includegraphics[width=7cm]{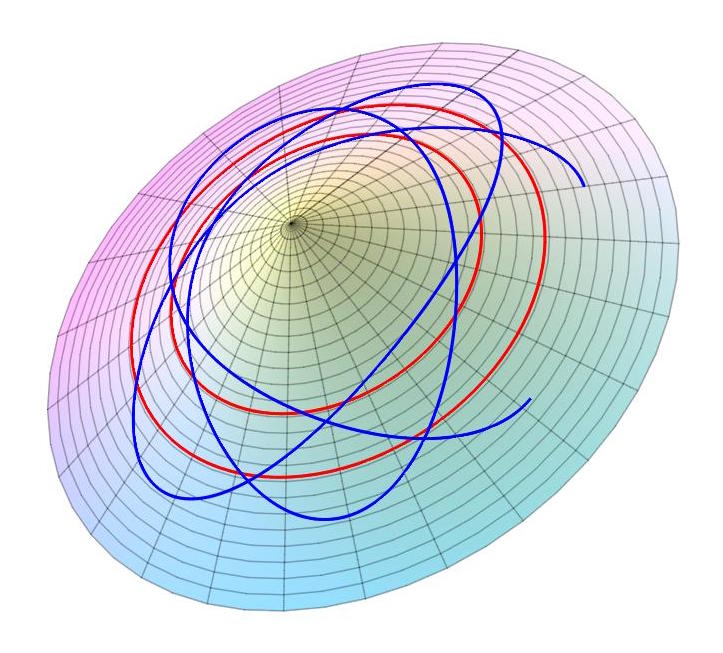} \hspace{1.4cm} \includegraphics[width=6.2cm]{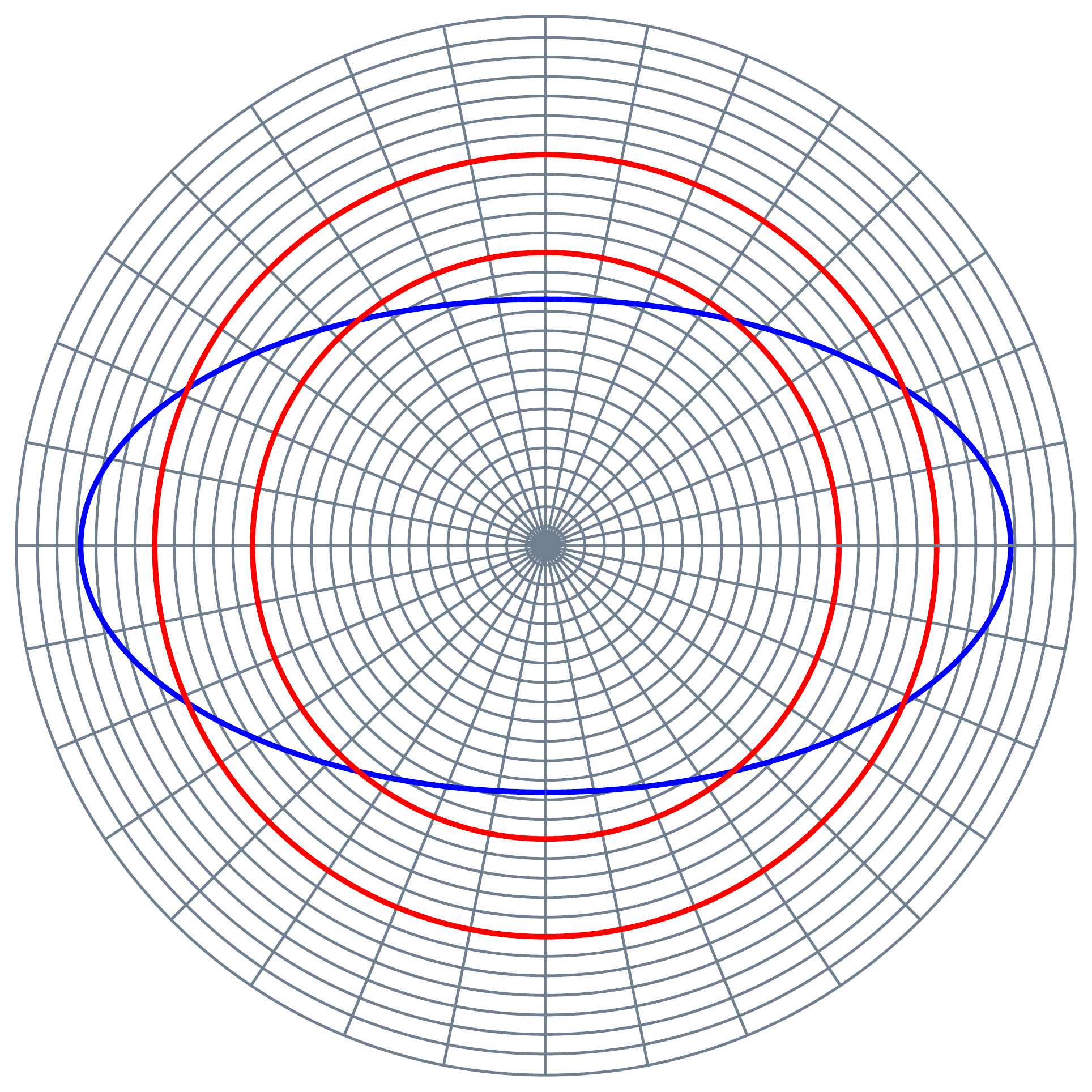}}
\caption{The blue curves are fluid streamlines for BTZ in the conical acoustic coordinates of section \ref{sect:ConicalAcoustic}.  On the left, a streamline on a cone with $c = 0.88$.
On the right, a streamline on a plane with $c = 1$.  The red circles indicate the inner and outer horizons.\label{fig:ConicalBTZ}}
\end{figure}

The equation ${dr \over d\phi} = {v^r \over v^\phi}$ can be integrated to find the fluid streamlines, which up to a shift of $\phi$ are given by
\be
r^2 = {r_{\rm max}^2 r_{\rm min}^2 \over r_{\rm max}^2 \cos^2(c \phi) + r_{\rm min}^2 \sin^2(c \phi)}
\ee
In general the fluid flows along a precessing ellipse on a cone, as shown in the left panel of Fig.\ \ref{fig:ConicalBTZ}.  When $c = 1$ the flow is along a closed ellipse on a plane,
as shown in the right panel of Fig.\ \ref{fig:ConicalBTZ}.

\subsection{Cosmological acoustic metric for $J = 0$\label{sect:BTZJ=0}}

Cosmological coordinates lead to a different fluid description of the BTZ black hole with some instructive new features.
Here we develop this for a non-rotating black hole.

To describe a non-rotating black hole we set $r_- = 0$ so that
\be
\label{J=0}
ds^2_{\rm BTZ} = - {r^2 - r_+^2 \over \ell^2} dt^2 + {\ell^2 \over r^2 - r_+^2} dr^2 + r^2 d\phi^2
\ee
To motivate the introduction of cosmological coordinates we reduce along $\phi$ and consider the $rt$ plane, described by
\be
\label{AdS2Rindler}
ds^2_{{\rm AdS}_2} = - {r^2 - r_+^2 \over \ell^2} dt^2 + {\ell^2 \over r^2 - r_+^2} dr^2
\ee
In fact this is AdS${}_2$ in Rindler coordinates \cite{Achucarro:1993fd}, which can also be described as an open FRW cosmology with metric
\be
\label{AdS2FRW}
ds^2_{{\rm AdS}_2} = - d\tau^2 + \ell^2 \cos^2 {\tau \over \ell} \, d\chi^2
\ee
To see that both (\ref{AdS2Rindler}) and (\ref{AdS2FRW}) describe patches of AdS${}_2$, recall that AdS${}_2$ is a
hyperboloid $-U^2-V^2+X^2 = -\ell^2$ in ${\mathbb R}^{2,1}$ with metric $-dU^2-dV^2+dX^2$.  We can introduce coordinates on the
hyperboloid in two ways.
\bea
\nonumber
U &=& {\ell r \over r_+} \enskip = \enskip \ell \cos {\tau \over \ell} \cosh \chi \\
\label{UVX}
V &=& \ell \, \Big({r^2 \over r_+^2} - 1\Big)^{1/2} \sinh {r_+ t \over \ell^2} \enskip = \enskip \ell \sin {\tau \over \ell} \\
\nonumber
X &=& \ell \, \Big({r^2 \over r_+^2} - 1\Big)^{1/2} \cosh {r_+ t \over \ell^2} \enskip = \enskip \ell \cos {\tau \over \ell} \sinh \chi
\eea
The induced metrics are (\ref{AdS2Rindler}) and (\ref{AdS2FRW}) respectively.

Motivated by this we apply the change of coordinates (\ref{UVX}) to the BTZ metric (\ref{J=0}), obtaining
\be
\label{CosmologicalBTZ}
ds^2_{\rm BTZ} = - d\tau^2 + \ell^2 \cos^2 {\tau \over \ell} \Big(d\chi^2 + {r_+^2 \over \ell^2} \cosh^2 \chi \, d\phi^2\Big)
\ee
Here $- {\pi \ell \over 2} < \tau < {\pi \ell \over 2}$, $-\infty < \chi < \infty$ and $\phi \approx \phi + 2 \pi$.
These cosmological coordinates cover the shaded region of the BTZ Penrose diagram shown in Fig.\ \ref{fig:BTZ}.

\begin{figure}
\centerline{\includegraphics[width=6cm]{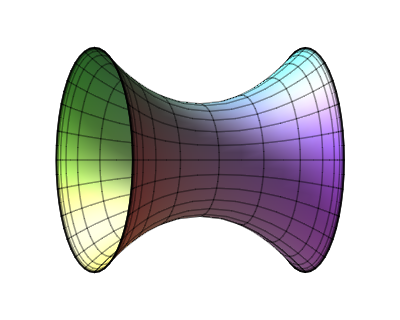}}
\caption{The wormhole geometry, conformally compactified and embedded in ${\mathbb R}^3$.\label{fig:wormhole}}
\end{figure}

As we now show, this metric can be interpreted as describing a comoving fluid on a cosmological Einstein-Rosen bridge.  First note that the spatial slices
in (\ref{CosmologicalBTZ}) are cosmological wormholes.  That is, the spatial metric has the form
\be
ds^2_{\rm spatial} =  a^2(\tau) ds^2_{\rm wormhole}
\ee
where the wormhole metric and scale factor are
\bea
\label{wormhole}
&& ds^2_{\rm wormhole} =  d\chi^2 + {r_+^2 \over \ell^2} \cosh^2 \chi \, d\phi^2 \\
\nonumber
&& a(\tau) =  \ell \cos {\tau \over \ell}
\eea
The wormhole geometry approaches a hyperbolic plane as $\chi \rightarrow \pm \infty$.\footnote{Up to a shift in $\chi$ the metric approaches
$ds^2_{\mathbb H^2} = d\chi^2 + \sinh^2 \chi \, d\phi^2$.}  Thus it describes a pair of Poincar\'e disks connected by a throat.\footnote{As $r_+ \rightarrow 0$
the throat pinches off, so the $M = 0$ geometry is singular.  Empty AdS corresponds to $M = -1$ \cite{Banados:1992wn}.}  The wormhole is illustrated in Fig.\ \ref{fig:wormhole}.
To obtain this figure we conformally compactified the wormhole geometry, defining
\be
\widetilde{ds}^2_{\rm wormhole} = {1 \over 4 \cosh^4 \chi / 2} ds^2_{\rm wormhole}
\ee
and embedded the compactified surface in ${\mathbb R}^3$.

To complete the fluid description note that the fluid velocity vanishes,
\be
v^\chi = v^\phi = 0
\ee
and the speed of light $c = 1$.  The vanishing fluid velocity just means the fluid is comoving with the spatial coordinates as the wormhole expands and contracts.  As in (\ref{FRW2}) we could
introduce $R = a(\tau) \chi$, $\Phi = a(\tau) \phi$ to make this motion explicit.  But unlike the case of flat FRW, in the present context this change of coordinates would
leave us with a time-dependent spatial metric.  So it seems more natural to leave the metric in the comoving form (\ref{CosmologicalBTZ}).

\subsection{Cosmological spinning BTZ\label{sect:Spinning_BTZ}}

Finally we introduce cosmological coordinates for a spinning BTZ black hole.  The construction can be motivated by recalling that BTZ is a quotient of AdS${}_3$
by a particular element of the $SO(2,2)$ isometry group \cite{Banados:1992gq}.  So we begin by introducing coordinates on AdS${}_3$, regarded
as a hyperboloid
\be
- U^2 - V^2 + X^2 + Y^2 = - \ell^2
\ee
embedded in ${\mathbb R}^{2,2}$ with metric $-dU^2-dV^2+dX^2+dY^2$.  A convenient set of coordinates is
\bea
\nonumber
&& U = \ell \cos {\tau \over \ell} \cosh {r_+ \phi \over \ell} \\
&& V = \ell \sin {\tau \over \ell} \cosh {r_+ \chi + r_- \phi \over \ell} \\
\nonumber
&& X = - \ell \sin {\tau \over \ell} \sinh {r_+ \chi + r_- \phi \over \ell} \\
\nonumber
&& Y = \ell \cos {\tau \over \ell} \sinh {r_+ \phi \over \ell}
\eea
These coordinates are induced by starting from a reference point $(\ell,0,0,0)$ and (i) rotating by $\tau / \ell$ in the $UV$ plane, (ii) boosting by $r_+ \phi / \ell$ in the $UY$
plane, and (iii) boosting by $-(r_+ \chi + r_- \phi)/\ell$ in the $VX$ plane.\footnote{This is set up to make the description of the spinning black hole as simple as possible.  The
coordinates (\ref{UVX}) we introduced in the spinless case are induced by a different sequence of $SO(2,2)$ transformations.}  In AdS${}_3$ the coordinate $\phi$ has infinite range.
To make a BTZ black hole the appropriate quotient is to identify $\phi$ with $\phi + 2 \pi$.  This gives
\bea
\label{Kasner}
&& ds^2_{\rm BTZ} = -d\tau^2 + \sin^2{\tau \over \ell} \, \big(r_+d\chi + r_-d\phi\big)^2 + r_+^2 \cos^2{\tau \over \ell} d\phi^2 \\
\nonumber
&& \hspace{1cm} 0 < \tau < {\pi \ell \over 2},\quad -\infty < \chi < \infty,\quad \phi \approx \phi + 2 \pi
\eea
Rather amusingly this describes a cosmology with two scale factors whose spatial slices are infinite flat cylinders.

To simplify the description it's convenient to reparametrize
\be
\phi \rightarrow \phi - q(\tau) \chi
\ee
where
\be
q(\tau) = {r_+ r_- \sin^2 {\tau \over \ell} \over r_-^2 + \big(r_+^2 - r_-^2\big) \cos^2 {\tau \over \ell}}
\ee
is chosen to make the spatial metric diagonal.  This brings the metric to the form (overdot = ${d \over d \tau}$)
\be
ds^2_{\rm BTZ} = -d\tau^2 + a_\chi^2 d\chi^2 + a_\phi^2 \big(d\phi - \dot{q} \chi d\tau\big)^2
\ee
where the scale factors are
\bea
a_\chi^2 & = & {r_+^4 \cos^2{\tau \over \ell} \sin^2{\tau \over \ell} \over r_-^2 + \big(r_+^2 - r_-^2\big) \cos^2 {\tau \over \ell}} \\[3pt]
\nonumber
a_\phi^2 & = & r_-^2 + \big(r_+^2 - r_-^2\big) \cos^2 {\tau \over \ell}
\eea
From this we can read off the fluid description.  The spatial geometry is an infinite cylinder.
\be
ds^2_{\rm spatial} = a_\chi^2 d\chi^2 + a_\phi^2 d\phi^2
\ee
The fluid is not quite comoving with the cylinder coordinates, rather it has a peculiar velocity
\be
v^\chi = 0, \qquad v^\phi = \dot{q} \chi
\ee
The speed of light $c = 1$.

Note that the radius of the spatial cylinder decreases monotonically from $r_+$ at $\tau = 0$ to $r_-$ at $\tau = \pi \ell / 2$, so these coordinates cover region {\sf II}
of the Penrose diagram shown in Fig.\ \ref{fig:Spinning_BTZ}.  One can send $r_- \rightarrow 0$ to describe a non-rotating black hole.  In this limit these coordinates
cover the future interior of the black hole, that is, the region $0 < r < r_+$.

\section{Conclusions\label{sect:conclusions}}

In this paper we developed fluid descriptions for various black hole geometries.  A fluid description, or equivalently an ADM decomposition, lacks manifest covariance and
is tied to a particular choice of coordinates.  Our goal was not to identify coordinate systems which led to realistic fluid flows that could be realized in the laboratory.  Rather we
developed fluid descriptions that helped us to understand and visualize properties of the spacetime geometry.  Thus our approach should be distinguished from much work in the
literature which is based on obtaining the effective Lorentzian geometry that governs small perturbations in realistic non-relativistic fluids.

An advantage of our approach is that, as a restatement of the ADM decomposition, it applies to any solution to general relativity and can be used to obtain useful intuition about the solution.
For example the fluid description of the Kerr metric in Doran coordinates captures the intuitive notion of space rotating as it is pulled in to a spinning black hole.  For BTZ we developed
several fluid descriptions which captured different aspects of the geometry.  Conical acoustic coordinates have simple static spatial slices, and the fluid flow captures the property that
both the AdS boundary and the BTZ singularity are repulsive.  Cosmological coordinates for $J = 0$ capture a complete spatial slice of the geometry, including the wormhole which the other
coordinate systems miss.  Finally cosmological coordinates for a spinning BTZ black hole provide a simple description of the region between the inner and outer horizon, in a form that
resembles a toy cosmology.

It would be interesting to explore the insights that can be gained by developing fluid descriptions for other solutions to general relativity.  It would also be interesting to develop a better physical
understanding of the space-fluid itself.  Can other fluid parameters be usefully introduced, beyond the fluid velocity and speed of light?

\bigskip\bigskip
\goodbreak
\centerline{\bf Acknowledgements}
\noindent
DK is grateful to Gary Gibbons for sparking his interest in this subject and to Cheryl Kang, Nazmul Islam and Max Brodheim for early collaboration.
The work of DK and AvG is supported by U.S.\ National Science Foundation grant PHY-1519705.  

\appendix
\section{Fluid interpretation and geodesic flow\label{geodesic}}

Consider the metric
\be
\label{AcousticMetric2}
ds^2 = - c^2 dt^2 + h_{ij} (dx^i - v^i dt) (dx^j - v^j dt)
\ee
which leads to the null cones
\be
\label{NullCones2}
h_{ij} \Big({dx^i \over dt} - v^i\Big) \Big({dx^j \over dt} - v^j\Big) = c^2
\ee
We generally assign $t$ units of length, so the velocities $v^i$ and $c$ are dimensionless.

In the fluid picture we view space as moving with coordinate velocity $v^i$ and we imagine that light rays move with speed $c$ -- meaning proper distance per unit coordinate time -- relative to the moving fluid.  The coordinate velocity of such
light rays ${dx^i \over dt}$ satisfies (\ref{NullCones2}).  An easy way to see this is to note that locally one can make a Galilean transformation $dx^{\prime i} = dx^i - v^i dt$, $dt' = dt$ to a frame
in which the fluid is at rest, and in this frame the condition (\ref{NullCones2}) becomes $h_{ij} {dx^{\prime i} \over dt'} {dx^{\prime j} \over dt'} = c^2$.  So the fluid picture leads to the null cones
(\ref{NullCones2}) associated with the metric (\ref{AcousticMetric2}).

It's worth noting that, as pointed out in \cite{Barcelo:2005fc}, the fluid moves along geodesics if and only if $c$ is independent of position.  To see this note that proper time along a trajectory
$x^i(t)$ in the metric (\ref{AcousticMetric}) is
\be
\tau = \int dt \sqrt{c^2 - h_{ij} (\dot{x}^i - v^i) (\dot{x}^j - v^j)}
\ee
where an overdot denotes a time derivative.  Varying with respect to the
trajectory and making the ansatz $\dot{x}^i = v^i$ one finds that the geodesic
equation is satisfied if and only if $\partial_i c = 0$.  That is, the fluid moves
along geodesics if and only if the speed of light is independent of position.  Note that this is
the case for all geometries considered in this paper.

\section{Doran coordinates\label{BL-D}}

The Kerr metric is frequently described in Boyer - Lindquist coordinates
\bea
\label{BL}
ds^2 & = & -dt^2 + {2 m r \over r^2 + a^2 \cos^2 \theta} \left(dt - a \sin^2 \theta d\phi\right)^2 \\[5pt]
\nonumber
& & + {r^2 + a^2 \cos^2 \theta \over r^2 + a^2 - 2 m r} dr^2 + \left(r^2 + a^2 \cos^2 \theta\right) d\theta^2 + \left(r^2 + a^2\right) \sin^2 \theta d\phi^2
\eea
The metric is invariant under shifts of $t$ and $\phi$.  To pass to Doran coordinates we translate these coordinates by an amount which depends on $r$.
\bea
\nonumber
&& t \rightarrow t + f(r) \\
&& \phi \rightarrow \phi + g(r)
\eea
With
\bea
\nonumber
&& f' = - {\sqrt{2mr(r^2 + a^2)} \over r^2 + a^2 - 2 m r} \\[5pt]
&& g' = - {a \over r^2 + a^2 - 2 m r} \sqrt{2 m r \over r^2 + a^2}
\eea
we obtain
\bea
\nonumber
ds^2 & = & - dt^2 + {r^2 + a^2 \cos^2 \theta \over r^2 + a^2} \left[dr + {\sqrt{2 m r (r^2 + a^2)} \over r^2 + a^2 \cos^2 \theta} \left(dt - a \sin^2 \theta d\phi\right)\right]^2 \\[3pt]
& & + (r^2 + a^2 \cos^2 \theta) d\theta^2 + (r^2 + a^2) \sin^2\theta d\phi^2
\eea
This is the Kerr metric in Doran coordinates.


\providecommand{\href}[2]{#2}\begingroup\raggedright\endgroup

\end{document}